\documentclass[conference]{IEEEtran}
\IEEEoverridecommandlockouts
\usepackage{cite}
\usepackage{amsmath,amssymb,amsfonts}
\usepackage{bm}
\usepackage{algorithmic}
\usepackage{graphicx}
\usepackage{textcomp}
\usepackage{xcolor}
\usepackage[outdir=Figs/]{epstopdf}
\usepackage[ruled,vlined]{algorithm2e}
\usepackage{subcaption}
\usepackage[keeplastbox]{flushend}
\usepackage[skip=2pt,font=small]{caption}

\def\BibTeX{{\rm B\kern-.05em{\sc i\kern-.025em b}\kern-.08em
		T\kern-.1667em\lower.7ex\hbox{E}\kern-.125emX}}
\begin{document}
	
	\title{Deep Contextual Bandits for Fast Initial Access in mmWave Based User-Centric Ultra-Dense Networks}
	
	\author{
		\IEEEauthorblockN{Insaf Ismath$^\dagger$, K.B.Shashika Manosha$^\ddagger$, Samad Ali$^\dagger$, Nandana Rajatheva$^\dagger$, Matti Latva-aho$^\dagger$}
		\IEEEauthorblockA{\textit{$^{\dagger \ddagger}$Center for Wireless Communication, University of Oulu, Oulu, Finland} \\
			\\
			\{insaf.ismath, samad.ali, nandana.rajatheva, matti.latva-aho\}@oulu.fi$^\dagger$, manoshadt@gmail.com$^\ddagger$}
		\thanks{The authors thank Mr. Tarun Chawla from Remcom for extending technical support and expertise for this work.}
		\thanks{This work is supported in part by the Academy of Finland 6Genesis Flagship (grant 318927)}
	}
	\maketitle
	
	\begin{abstract}

		Millimeter wave (mmWave) based multiple-input multiple-output (MIMO) capable user-centric (UC) ultra-dense (UD) networks are suggested to facilitate high throughput requirements of future networks.
		Due to the high blockage susceptibility of mmWave, the connections may drop frequently. 
		Hence efficient and fast beam management in initial access (IA) is essential. 
		Current cellular systems use beam sweeping based IA mechanisms.
		UC UD concept requires all of its access points (APs) to perform IA.
		This leads to a shortage of orthogonal radio resources.
		Nonorthogonal resource allocation causes interference which leads to a higher misdetection probability.
		In this paper, we propose a novel deep contextual bandit (DCB) based approach to perform fast and efficient IA in mmWave based UC UD networks. 
		The DCB model uses one reference signal from the user to predict the IA beam.
		The reduced use of reference signals improves beam discovery delay and relaxes the requirement for radio resources.
		Ray-tracing and stochastic channel model-based simulations show that the suggested system outperforms its beam sweeping counterpart in terms of probability of beam misdetection and beam discovery delay in mmWave based UC UD networks.

	\end{abstract}

	\section{Introduction}
	To support futuristic applications like virtual reality and autonomous driving, wireless communication systems of the future require at least a 1000-fold increase of throughput\cite{Qualcomm2013}.
Over the past cellular generations, increased throughput requirements have been met by network densification; the number of access points (APs) deployed in a unit area is increased\cite{Andrews2015}.
%Cell sizes are reduced and compared to the cells in the preceding generation.
However, it is suggested that network densification is reaching its limit due to the increased inter-cell interference and uncoordinated AP transmissions of the current systems\cite{Interdonato2019, Andrews2015}.
The user-centric (UC) network topology is a novel concept proposed as an alternative to the cellular system to facilitate the massive throughput requirements of future networks\cite{Interdonato2019, Buzzi2017}.
The UC topology includes an ultra-dense (UD) AP distribution, massive multiple-input multiple-output (MIMO) capable APs, and a cooperative user serving system, to meet the demands of the future. 
The users in UC systems would be simultaneously served by a subset of available APs to improves throughput and reliability by creating diversity. 
Current 5th generation (5G) and future wireless communication systems are set to use the millimeter wave (mmWave) and sub-terahertz (THz) bands for communications to meet the throughput requirements.
Nevertheless, mmWave and THz channels have a higher susceptibility to blockages due to the high penetration losses it experiences compared to lower frequency bands\cite{Jain2019}. 

\begin{figure}[t]
	\centering
	\includegraphics[width=0.9\linewidth]{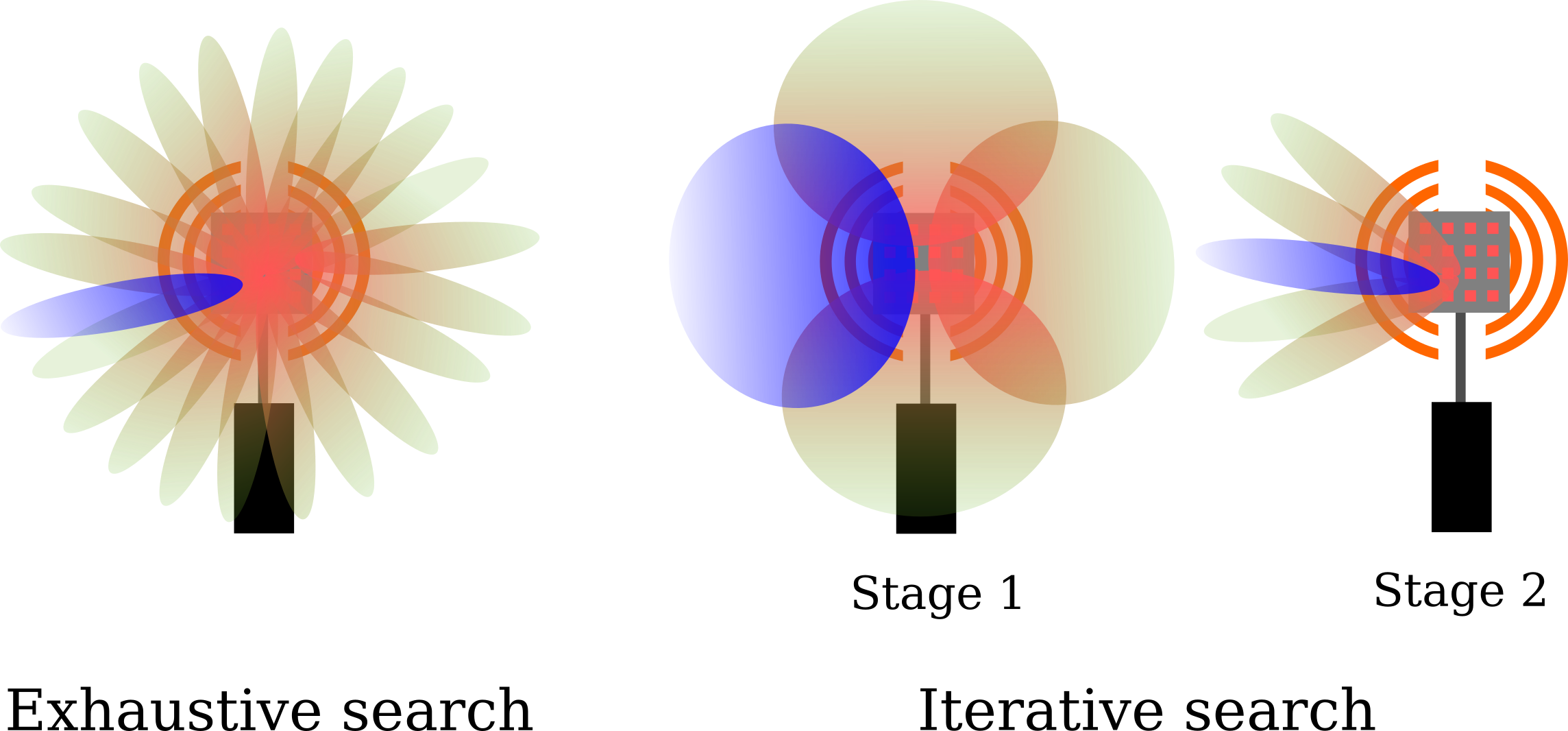}
	\caption{Beam transmissions in the iterative, and exhaustive search approaches.}
	\label{fig:iaapproches}
\end{figure}

The inevitable loss of connections in mmWave channels calls for fast and efficient beam management tools to maintain reliable wireless communications. 
The initial access (IA) process is responsible for establishing the connection between the APs and the users.
The beam management in the IA process selects the best beam from a predefined quantized beam codebook for an AP to serve a user.
IA systems deployed in current and past cellular systems depend on exhaustive or iterative search based beam sweeping mechanisms.
The exhaustive search approach conducts a full-beam sweep by transmitting all the beam in the codebook one after the other. 
Users select the beam providing the best signal-to-noise ratio. 
Iterative search is a two-stage beam sweeping strategy where at stage one, APs transmit a set of wider beams to narrow down the search area.
Subsequently, in stage two, a beam sweep is carried out within the identified area to fine-tune the beam selection.
Fig. \ref{fig:iaapproches} presents an abstract idea of how these beams are transmitted in iterative and exhaustive beam search. 
IA system in 5G employs an iterative search based approach\cite{3GPP2018}.
Beams are transmitted using orthogonal resources to avoid interference that would result in beam misdetection at the user.

Authors in \cite{Giordani2016} discuss and compared IA approaches suggested for 5G networks such as iterative, and exhaustive search.
The authors of \cite{Perera2020a} propose an algorithm that dynamically scales the resources allocated to beam sweeping directions to optimize the beam management of the IA procedure in 5G mmWave networks. 
Recently, machine learning (ML) based approaches have been explored as alternatives to replace conventionally non-ML bases systems in communication systems \cite{ali2020white, Mismar2020, Cousik2020}.
Work in \cite{ali2020white} presents a discussion on possible implementations of ML in future wireless communication systems to meet the requirements of the network.
Authors of \cite{ali2020white} suggest that ML will be vital in accommodating the increasing demand for connectivity.
The work in \cite{Cousik2020} presents an IA algorithm called DeepIA which uses a deep neural network (DNN) for faster prediction of the IA beam.
Instead of using all the beams in the codebook, DeepIA proposes to use only a subset of beams for beam sweep.
SNR reports from the users are used as the input for the DNN that predicts the IA beam.
The DeepIA algorithm improves on the two-stage iterative beam sweeping strategy by substituting the second stage with a DNN. 
However, the beam discovery delay caused multiple message transfers: beam transmissions and user reports, by beam sweeping based methods, can not be reduced.  
Authors in \cite{Mismar2020} present a solution that simultaneously solves beamforming, power allocation, and interference management based on deep reinforcement learning for cellular systems.
The algorithm given in \cite{Mismar2020} assumes access to real time knowledge of interference and received power experienced at the users. 
The implementation of \cite{Mismar2020} would require high fronthaul and backhaul capacities which would lead to a lack of feasibility in UC UD networks. 

The motivation for this work comes from trying to answer the following two questions regarding IA beam management in MIMO enabled UC UD networks based on mmWave. 
a) Does the system have enough orthogonal resources to support beam sweeping of the scale required by the MIMO capable UC UD networks?
b) Can the beam discovery delay be reduced by reducing the number of message transfers needed for IA beam identification?
Based on the discussion above, this work proposes a novel deep contextual bandit (DCB) based approach to perform fast and efficient IA for mmWave based UC UD networks. 
The proposed approach reduces the number of messages transferred between the APs and the users to lower the beam discovery delay and radio interference.
%The reduced redundancy caused by the lack of additional reference signals is compensated using machine learning methods.
The DCB model in the proposed approach is trained to predict the beamforming vector using only one reference signal from the user.
The performance of the suggested IA scheme is evaluated using stochastic and ray-tracing based methods.
Simulation results show that the proposed system is capable of maintaining beam misdetection probability close to zero for the considered network topology.

The rest of this paper is organized as follows. 
Section \ref{sec:sys_model} explains the system model used in our work.
Section \ref{sec:IA5G} provides a quick primer on the beam sweeping based IA scheme used in 5G and how it could be extended to the UC UD case.
Section \ref{sec:dcbia} introduces the DCB ML model and formally presents the DCB based IA scheme proposed in this work.
Section \ref{sec:perf} presents the simulation model and numerical results, while this work is concluded in Section \ref{sec:conclution}.

\underline{Notations}: $(.)^T$, and $(.)^H$ denote transpose and Hermitian transpose, respectively.
% $diag\{x\}$ represents a diagonal matrix wherein the elements in the main diagonal are defined by x.
% $I_n$ and $O_n$ represents the $n\times n$ identity and zero matrix, respectively.
$\mathcal{R}(\mathbf x)$ and $\mathcal{I}(\mathbf x)$ represents the real and imaginary parts of $\mathbf x$, respectively.
 $||.||$, and $|\mathcal{X}|$ denotes the euclidean norm and the cardinality of the set $\mathcal{X}$, respectively.

	\section{System Model}
	\label{sec:sys_model}
%This section introduces the system model used in this work and presents the problem.
%\subsection{System Model}

\begin{figure}[t]
	\centering
	\includegraphics[width=0.8\linewidth]{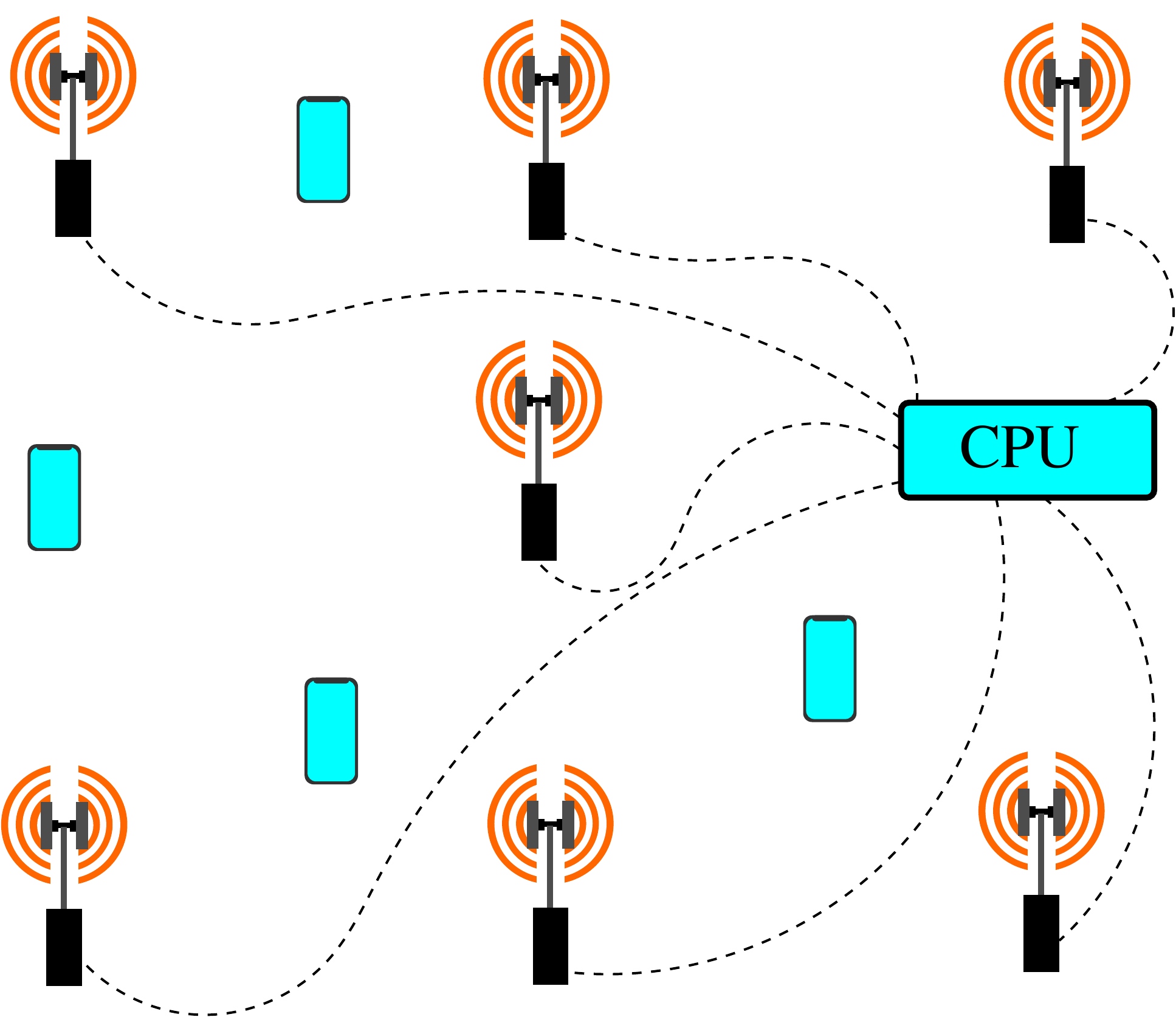}
	\caption{The system model of a user-centric ultra-dense network where a subset of the available APs would jointly serve a user\cite{Interdonato2019}. }
	\label{fig:systemmodel}
\end{figure}

Consider a UC UD network topology with a set $\mathcal{K}$ of ${K}$ single-antenna users are served by subsets of ${M}$ APs.
The user selects the best $M$ APs providing the best signal powers out of $\mathcal{M}$\cite{Interdonato2019}.
The entity managing the APs is referred to as the central processing unit (CPU).
The system model of a UC UD network is presented in Fig. \ref{fig:systemmodel}.
APs are equipped with $N$ element uniform rectangular planar antenna arrays (URPA).

The channel is modeled using a clustered mmWave model\cite{Akdeniz2014} with $J$ clusters.
Each cluster is generated with $L$ paths which are parameterized by path loss, fading, and array gain at the AP.
The channel from an user to an AP, $\mathbf{g} \in \mathbf{C}^{N\times 1}$, is presented as
\begin{equation}
\mathbf{g} = \frac{1}{\sqrt{L}}\sum_{j=1}^{J} \sum_{l=1}^{L} p_{j,l}h_{j,l}\mathbf{a}(\theta_{j,l},\phi_{j,l}),\label{eq:gkmn}
\end{equation}
where 
$p_{j,l}\in \mathbf{C}$ is the path loss,
$h_{j,l} \in \mathbf{C}$ is the small-scale fading gain, and 
$\mathbf{a}(\theta_{j,l},\phi_{j,l}) \in \mathbf{C}^{N\times 1}$ is the AP's array gain, where $\theta_{j,l}$ and $\phi_{j,l}$ are the azimuth and elevation angles of arrival, respectively.
Here, the subscript notations $j=1,\dots,J$ and $l=1,\dots,L$ represent cluster and path index, respectively. 
%The channel matrix $\mathbf{g}$ from the $k$th user to the $m$th AP is denoted as $\mathbf{g_{k,m}}$.
Channel reciprocity is assumed; all mmWave channels proposed in the frequency region 2 for 5G and future systems use time division duplexing\cite{FR2}.     
Hence the channel from the AP to the user is $\mathbf{g}^H$.

Beams for IA are chosen from a predefined beam codebook $\mathcal{C}$ and they are assumed to be implemented with a network of quantized phase shifters.
The $b$th entry of $\mathcal{C}$, i.e., $\mathbf{f_b}$, is given by
\begin{equation}
\mathbf{f_{b}} = \frac{1}{\sqrt{N}} \left[ e^{j\Theta_{b,1}} \; \dots \;e^{j\Theta_{b,N}}\right] ^T,
\end{equation}
where
$\Theta_{b,n}$ is the quantized phase shift corresponding to the $n$th antenna in the $b$th entry.

The mmWave channels generated using a commercial ray-tracing software called Wireless Insite\cite{REMCOM2017} allows performance analysis in a realistic environment. 
However, it is practically infeasible to study the effects of the time-varying nature of the channel on performance using a ray-tracing based approach. 
Hence, simulations based on stochastic channel models are also considered.

	\section{Extension of 5G IA to User-centric Systems}
	\label{sec:IA5G}
\begin{figure}[t]
	\centering
	\includegraphics[width=\linewidth]{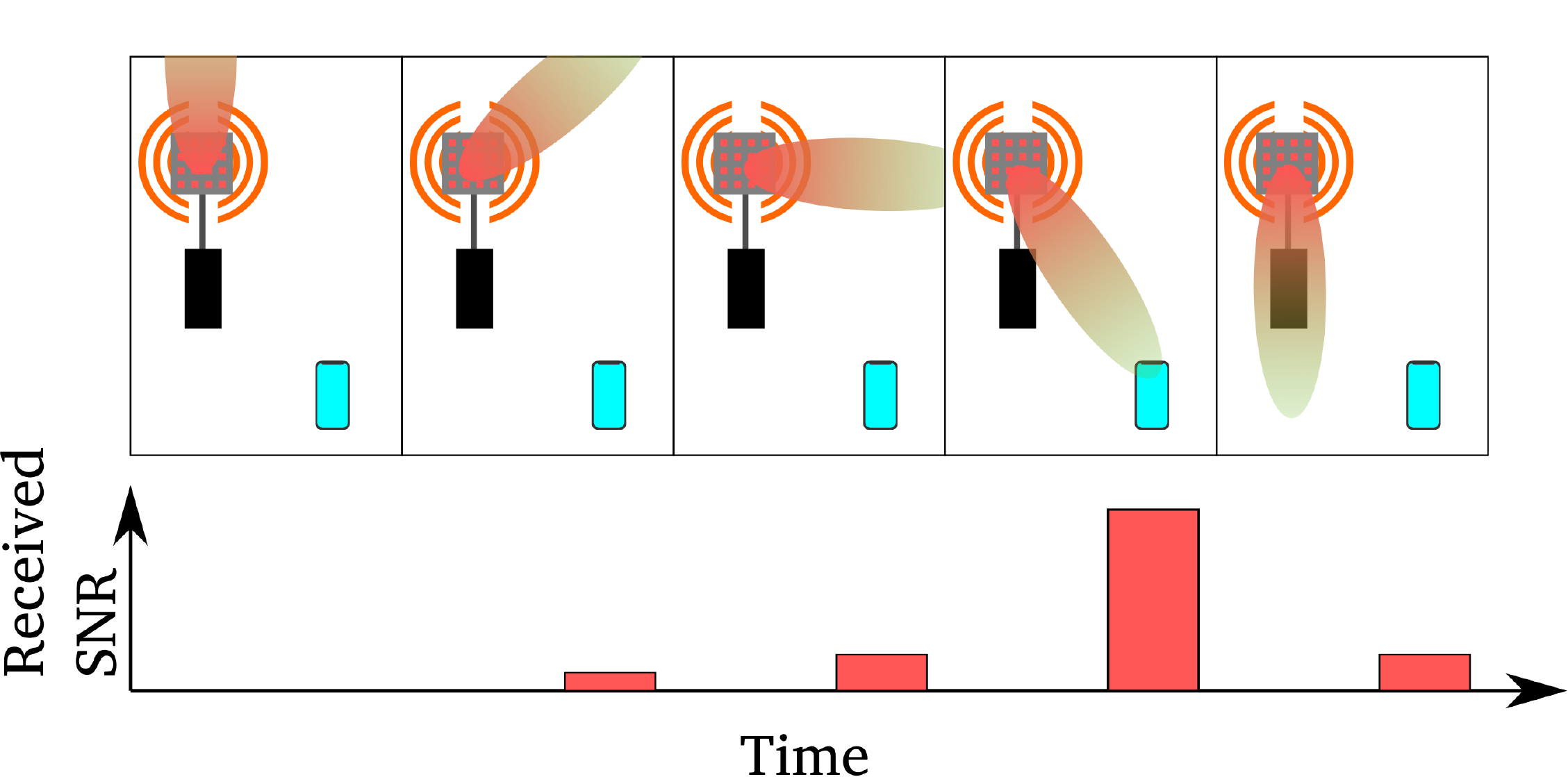}
	\caption{Iterative beam transmissions in 5G IA procedure over time.}
	\label{fig:5gbeammanagement}
\end{figure}

IA in 5G systems uses a beam management strategy based on an exhaustive search\cite{3GPP2018}. 
IA beams in the codebook are transmitted using synchronization signal blocks (SSBs) in a predetermined order over time as seen in Fig. \ref{fig:5gbeammanagement}.
Beams are evaluated at the user in terms of the signal-to-interference-plus-noise ratio (SINR).
The beam with the best SINR performance is reported back to the respective AP. 
The user then associates itself with the AP that provided the aforementioned best beam.

Applying the 5G IA approach to the single-user single-AP case is straightforward.
The goal of this problem is to maximize the SNR experienced at the user by appropriate selection of the IA beam.
The SNR at the user is expressed as
\begin{align}
\operatorname{SNR} =  \frac{P||\mathbf{g}^H\mathbf{f}||^2}{\sigma^2},
\end{align}
where 
$P$ is the total transmit power of the AP,
$\mathbf{g}^H$ is the channel from the AP to the user,
$\mathbf{f}$ is the beam chosen by the user to be served from the AP, and
$\sigma^2$ is the noise power at the user.
This case is similar to IA in traditional systems, and the maximization problem is presented as 
\begin{equation}
\begin{aligned}
\mathop{\operatorname {maximize} } _{\mathbf{f}} \quad & \operatorname {SNR}\\
%\textrm{s.t.}  y_{i}(w\phi(x_{i}+b))+\xi_{i}-1\\
\operatorname {subject \; to} \quad & \;\; \mathbf{f} \in \mathcal{C}.
\end{aligned}
\end{equation}

IA approach in 5G is extended to the multi-AP setting of the UC UD network architecture with some minor changes.
Since the UC concept allows multiple APs to cooperatively serve the users, each AP has to perform the IA procedure despite the high AP density of 40-200 APs per km$^ 2$\cite{Lin2018, Pan2019} .
After evaluating the beam SINRs, users have to select the best beam for each AP.
The user then associates itself with the best $M$ APs which provide the highest total SNIR.
%compared to selecting AP in the cellular case. 
The goal of this problem is to maximize the sum of SINR received at the user by selecting the best beam from each AP.
The total SINR received at $k$th user is presented as,
\begin{align}
\operatorname {SINR_k} =  \frac{\sum_{m =1}^{M}P_{m,k}||\mathbf{g}^H_\mathbf{{m,k}} \mathbf{f_{m,k}}||^2}{I_k+\sigma_k^2},
\end{align}
where
$P_{m,k}$ is the transmit power allocated to $k$th user at $m$th AP,
$\mathbf{g}^H_\mathbf{{m,k}}$ is the channel from the $m$th AP to the $k$th user, 
$\mathbf{f_{m,k}}$ is the beam chosen by the $m$th AP to serve the $k$th user, and 
$\sigma_k^2$ is the noise power at $k$th user.
%Beam sweeping in UC UD networks 
Beam transmissions use orthogonal SSBs to avoid interference.
%To perform IA in a UC UD network,  
Due to the high density of APs, UC UD networks need a higher number of resources.
The limited radio resources could be either reused frequently or shared using nonorthogonal allocation schemes\footnote{For the practical deployment of mmWave channels, specifications in \cite{3GPP2018} define a maximum of 64 SSBs.
Considering the high-density nature of the UC UD network, this number is inadequate for all the APs to perform interference free beam sweeping.
This lack of radio resources causes beam interference. }.
However, this causes interference which affects the beam misdetection probability.
We define the probability of a beam interfering with another beam at a user as $q_i$.
The interference power caused at $k$th user is $I_k$.
This problem is presented as
\begin{equation}
\begin{aligned}
\mathop {\operatorname {maximize} } _{\mathbf{f_{1,k}},\dots, \mathbf{f_{M,k}}} \quad & \operatorname {SINR_k}\\
%\textrm{s.t.}  y_{i}(w\phi(x_{i}+b))+\xi_{i}-1\\
\operatorname {subject \; to} \quad & \;\; \mathbf{f_{m,k}} \in \mathcal{C} \;\;\; \forall m.
\end{aligned}
\end{equation}

	\section{Deep Contextual Bandit Based Proposed Approach for IA}
	\label{sec:dcbia}
This section provides an introduction to DCB and introduces the main contribution of this work--the novel DCB based approach to perform IA in mmWave based UC UD networks.

\subsection{Primer On Deep Contextual Bandits}
\label{subsec:dcb-primer}
\begin{figure}[t]
	\centering
	\includegraphics[width=0.8\linewidth]{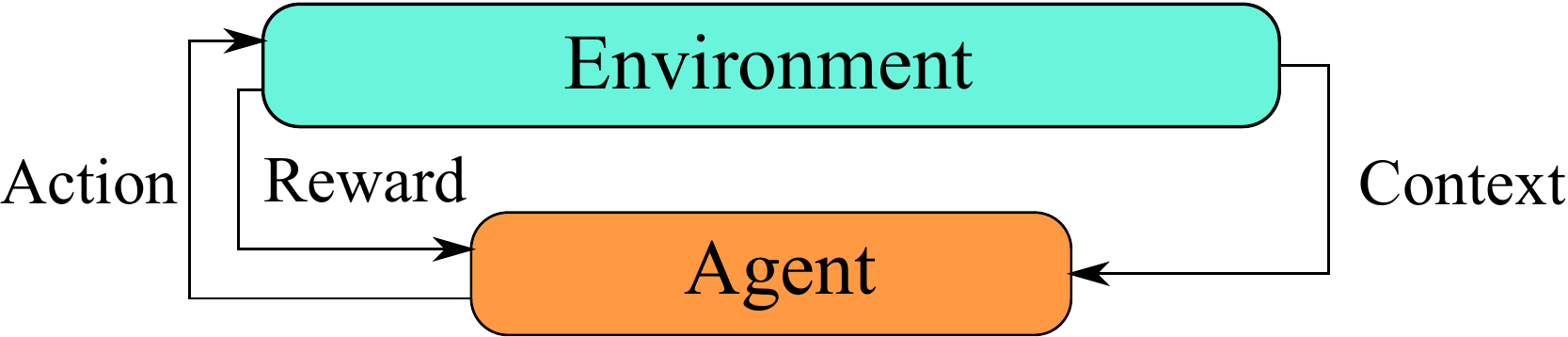}
	\caption{Abstract structure of the deep contextual bandit algorithm.}
	\label{fig:drlmodel}
\end{figure}

The architecture of a deep contextual bandit (DCB) machine learning model is presented in Fig. \ref{fig:drlmodel}.
The software entity which interacts with the environment is called an agent. 
The agents use actions ($a$) from a set action-space to interact with the environment.
The state of the environment is characterized by context $\mathcal{X}$. 
The agent is then provided with feedback on this interaction in the form of a reward $\gamma$.
The agent's goal is to select actions to maximize the received reward.  
Hence, the agent has to learn the mapping function that maps a context to an action that gives the maximum received reward.
This is done by building a model to predict the probability of receiving the highest reward for each action given the context; the model learns the probability set $\left\lbrace \operatorname{Pr}\left\lbrace \gamma|a, \mathcal{X}\right\rbrace\right\rbrace  \;\forall \gamma,a, \mathcal{X}$ \cite{Collier}.
For problems with small action-spaces, and discrete $\mathcal{X}$, the simplest model is a reward table which contains the $\gamma$ for each action-context pair.
Maintaining a reward table for problems with continuous $\mathcal{X}$ and large action-spaces is infeasible\cite{Mnih2015}.
Hence, the mapping function is approximated using a deep neural network (DNN).

DCB agents are deployed without prior training and they have to learn on the job. 
Hence deciding the compromise between exploitation and exploration policies is a dilemma in DCBs\cite{Collier}. 
The agent takes random actions to explore the environment under exploration to learn more about the environment. 
However, the agent could follow exploitation, i.e., greedy policy, and use its current knowledge to select actions to maximize rewards.
This work adopts a diminishing $\epsilon$-greedy policy where the agent follows exploration and exploitation policies with the probabilities of $\epsilon$ and 1-$\epsilon$, respectively\cite{Riquelme2017}.
Initially, $\epsilon$ is set to 1, and it is diminished by a factor of $\epsilon_{dec}$ at each training episode until it reaches a predefined minimum, i.e., $\epsilon_{min}$.
Defining $\epsilon_{min}$ is important since it allows the model to evolve with changing environmental conditions even after the initial learning period. 

The $\left( \mathcal{X}, a, \gamma\right) $ triplets are saved as experiences in the agent's memory.
At each training session, a certain number of experiences are chosen randomly and used as the training data.
This is called experience-replay and it promotes the convergence of the DCB model\cite{Riquelme2017}.

\subsection{Proposed Approach}

The proposed approach finds the beam for IA using just one reference signal from the user.
This reduction in the use of reference signals helps to lower the beam discovery delay.
It further improves on the efficiency by maintaining a low probability of beam misdetections.
Each AP will rely on the DCB agent deployed in them to identify the best IA beam.

Unlike 5G IA, the user initiates the IA procedure by broadcasting one reference signal.
This transmission is received by the nearby APs and it is used by the DCB agents to predict the IA beam.
%The reference signal is received by the $N$ antennas at the AP.
The received signal at the AP $\mathbf{y} \in \mathbf{C}^{N\times 1}$, is presented as
\begin{equation}
\mathbf{y} = \sum_{k \in \tilde{\mathcal{K}}}p_k\mathbf{g_{k}}\Omega_k^H + \mathbf{\omega},
\end{equation}
%where $\tilde{K}$ is the number of active users requesting IA, $\Omega_k$ is the reference signal transmitted by the $k^{th}$ user and $\omega_m$ is the $\mathcal{N}\times1$ complex vector containing noise at the receiver of $m^{th}$ AP. 
where %$\mathbf{y} = [y_{1} \;\dots \;y_{N}]^T$,
$\mathbf{\omega} = [\omega_{1} \;\dots \;\omega_{N}]^T$, 
$p_k$ is the transmitted power of $k$th user, and
$\mathbf{g_{k}}$ is the channel $\mathbf{g}$ from $k$th user to the AP.
Here
$\tilde{\mathcal{K}}\subseteq \mathcal{K}$ denotes the set of users requesting IA, and
%$\{y_{n}\}$ are the signal received by the $n$th antenna of the AP, and
$\{\omega_{n}\}$ are independent and identically distributed circularly-symmetric complex Gaussian noise at $n$th antenna of the URPA.
The noise has a mean of zero, and a variance of $\sigma^2$ per dimension.
The reference signal broadcasted by $k$th user is denoted as $\Omega_k$, and assume that $||\Omega_k||^2 = 1$ and $||\Omega_j^H\Omega_k||^2 = 0 \; \forall j\neq k$.
%and is a pilot transmitted as the reference signal by $k$th user.
Each AP follows a pilot matched channel estimation method \cite{Buzzi2017} to recover the channel matrix from $k$th user to the AP, and it is presented as
\begin{equation}\label{gkm_0}
\begin{aligned}
\mathbf{\tilde{g}_{k}} &= \mathbf{y}\Omega_k\\
&=p_k\mathbf{g_{k}}+ \mathbf{\omega}\Omega_k,
\end{aligned}
\end{equation}
where $\mathbf{\tilde{g}_{k}} \in \mathbf{C}^{N\times 1}$ is the estimated channel.
The real and the imaginary parts of the normalized $\mathbf{\tilde{g}_{k}}$, $\mathcal{R}\left(\mathbf{\tilde{g}_{k}}\right) \in \mathbf{R}^ {N\times1}$ and $\mathcal{I}\left(\mathbf{\tilde{g}_{k}}\right) \in \mathbf{R}^ {N\times1}$, respectively, are separated and then merged to create the input context $\mathcal{X} \in \mathbf{R}^ {2N\times1}$ and it is given by
\begin{equation}\label{context}
\mathcal{X}= \frac{1}{||\mathbf{\tilde{g}_{k}}||^ 2}\left[ \mathcal{R}\left(\mathbf{\tilde{g}_{k}}\right)^T , \mathcal{T}\left(\mathbf{\tilde{g}_{k}}\right)^T\right] ^T.
\end{equation}

The action-space of the DCB agent and the beam codebook $\mathcal{C}$ has a one-to-one mapping; each action corresponds to a beam in $\mathcal{C}$. 
Hence, the number of total actions equals $|\mathcal{C}|$.

After selecting an action, the AP uses the corresponding beam $\mathbf{f}$ for the downlink transmissions. 
The user receiving evaluates the downlink SNR and sends a report back to the AP.
The reward calculation for the DCB model is based on this SNR report.
Apart from $\mathbf{f}$, the SNR depends on many factors such as channel from the AP to $k$th user $\mathbf{h_k}\in \mathbf{C}^{1\times N}$, shadowing, and noise at the user.
The reward $\gamma$ is supposed to give feedback to the agent only about their actions.
Hence, the direct usage of the received SNR as the reward might confuse the agent.
Normalizing the SNR with $||\mathbf{h_k}||^2$ and AP transmit power can remove some of the dependencies.
However, $\mathbf{h_k}$ is not available at the AP.
We assume that the uplink and downlink transmissions occur within the channel coherence time, and therefore, the reciprocity still holds.
Hence, $\mathbf{h_k}$ is approximated using $\mathbf{\tilde{g}_{k}}^H$.
The reward metric $\gamma$ is presented as
\begin{equation}
\gamma = \frac{||\mathbf{h_{k}}\mathbf{f}||^2}{\sigma_k^2||\mathbf{\tilde{g}_{k}}^H||^2}\label{eq:norm_reward}.
\end{equation}

After saving this experience, the DCB model performs an experience-replay to train itself as explained in Section \ref{subsec:dcb-primer}.
Hence the IA process is concluded.
%Nevertheless, if the user is either not satisfied with the beam or if the beam didn't reach the user at all, they may initiate the IA process once again.
	
	\section{Performance Analysis}

	\begin{figure}[t]
	\centering
	\begin{subfigure}[t]{0.6\linewidth}
		\centering
		\includegraphics[width=\linewidth]{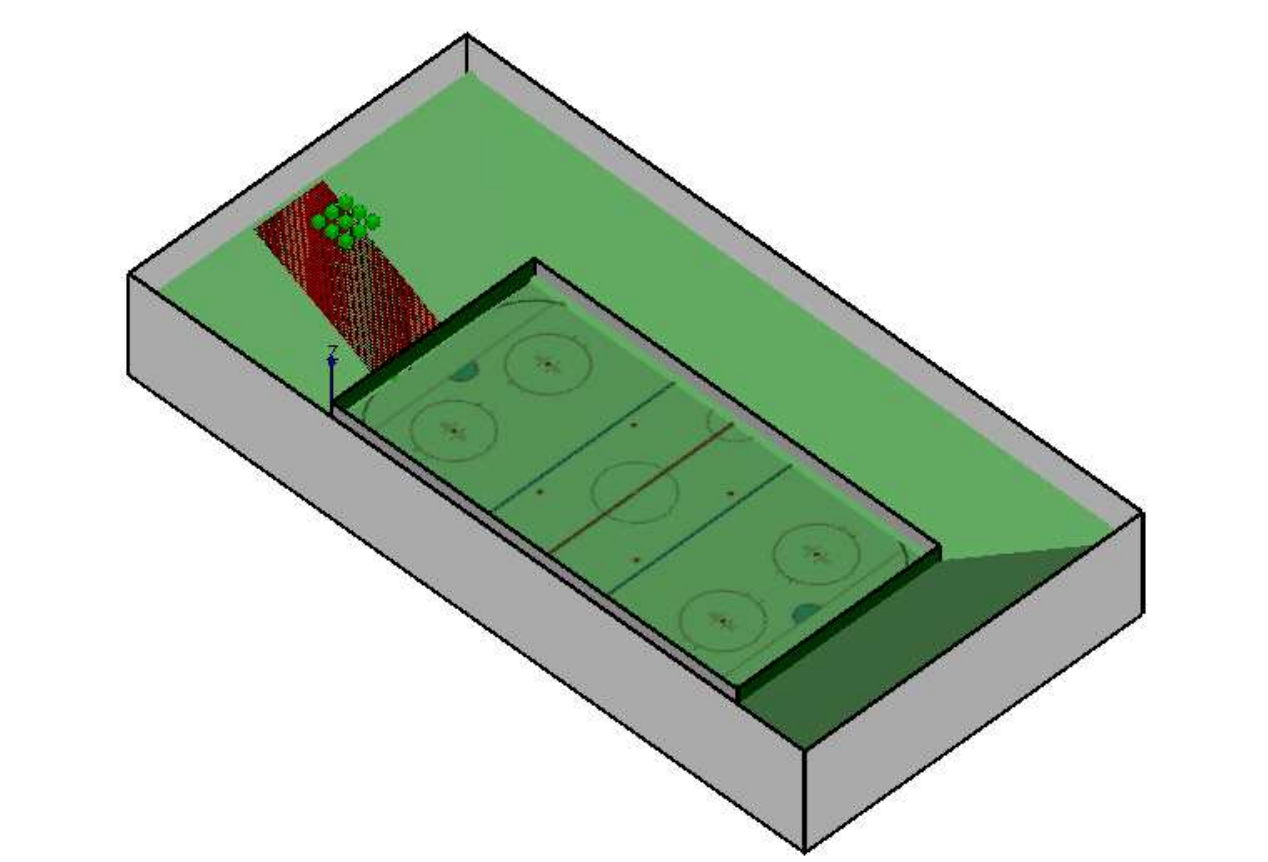}
		\caption{South-eastern view}
		\label{fig:simenvironment-se}
	\end{subfigure}%
	~     
	\begin{subfigure}[t]{0.4\linewidth}
		\centering
		\includegraphics[width=0.5\linewidth]{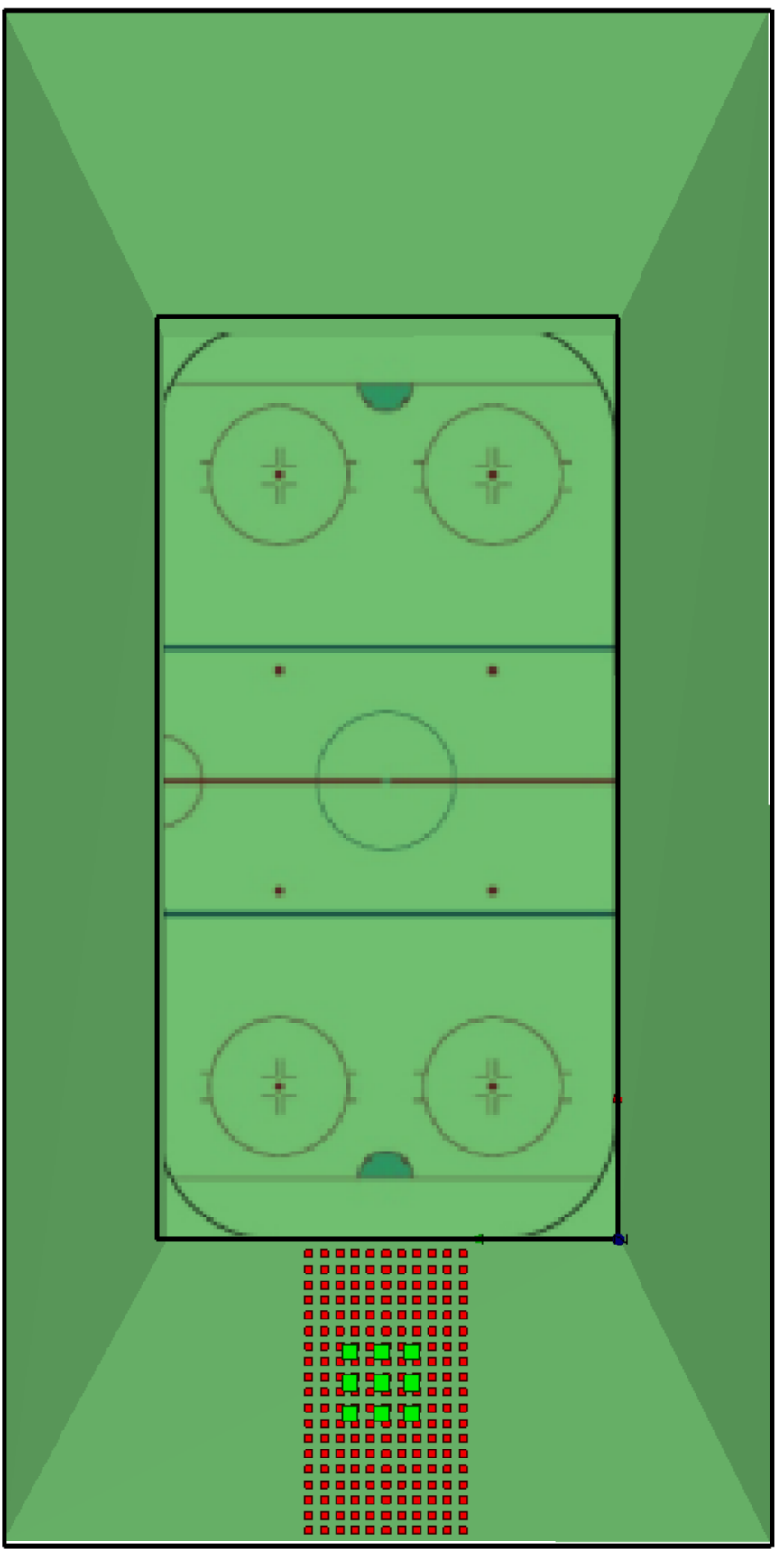}
		\caption{Top view}
		\label{fig:simenvironment-t}
	\end{subfigure}
	\caption{Simulation environment of the ray-tracing model.}
	\label{fig:simenvironment}
\end{figure}
	\label{sec:perf} 
This section presents the details about the simulations and demonstrates the numerical results.
\subsection{Simulation Model}
A carrier frequency of 28 GHz with a channel bandwidth of 15 MHz is considered.
Each AP is equipped with a 4 $\times$ 4 URPA. 
The beam codebook consists of 16 beams, i.e, $|\mathcal{C}| = 16$.
The maximum transmit power of the APs, and the users are set to 43 and 23 dBm, respectively.
It is assumed that every IA related message, i.e., reference signals, user reports, and beam transmissions, transferred between APs and users take 0.01 ms.
The processing time required for prediction is negligible compared to message transfer time.
We assume the probability of a beam transmitted in a beam sweep  causing interference to another beam is $q_i$.
Simulations are performed for different $q_i$ values.
The proposed approach relies on just one reference signal from the user for beam prediction. 
Here in order to detect the user reference signal at the APs, a minimum SNR threshold of 0 dB is considered.

Simulations are repeated independently for $10^ 6$ iterations to ensure the statistical significance of the results.
Performance is evaluated in terms of beam discovery delay and beam misdetection probability.
Beam discovery delay relates to the time required by the IA process to associate the user by providing an IA beam. 
Beam misdetection probability concerns with the quality of the beam choice; if the selected beam is unable to produce at least 95\% of the power delivered by the best beam choice, it is considered as a misdetection.

\begin{figure}[t]
	\centering
	\includegraphics[width=\linewidth]{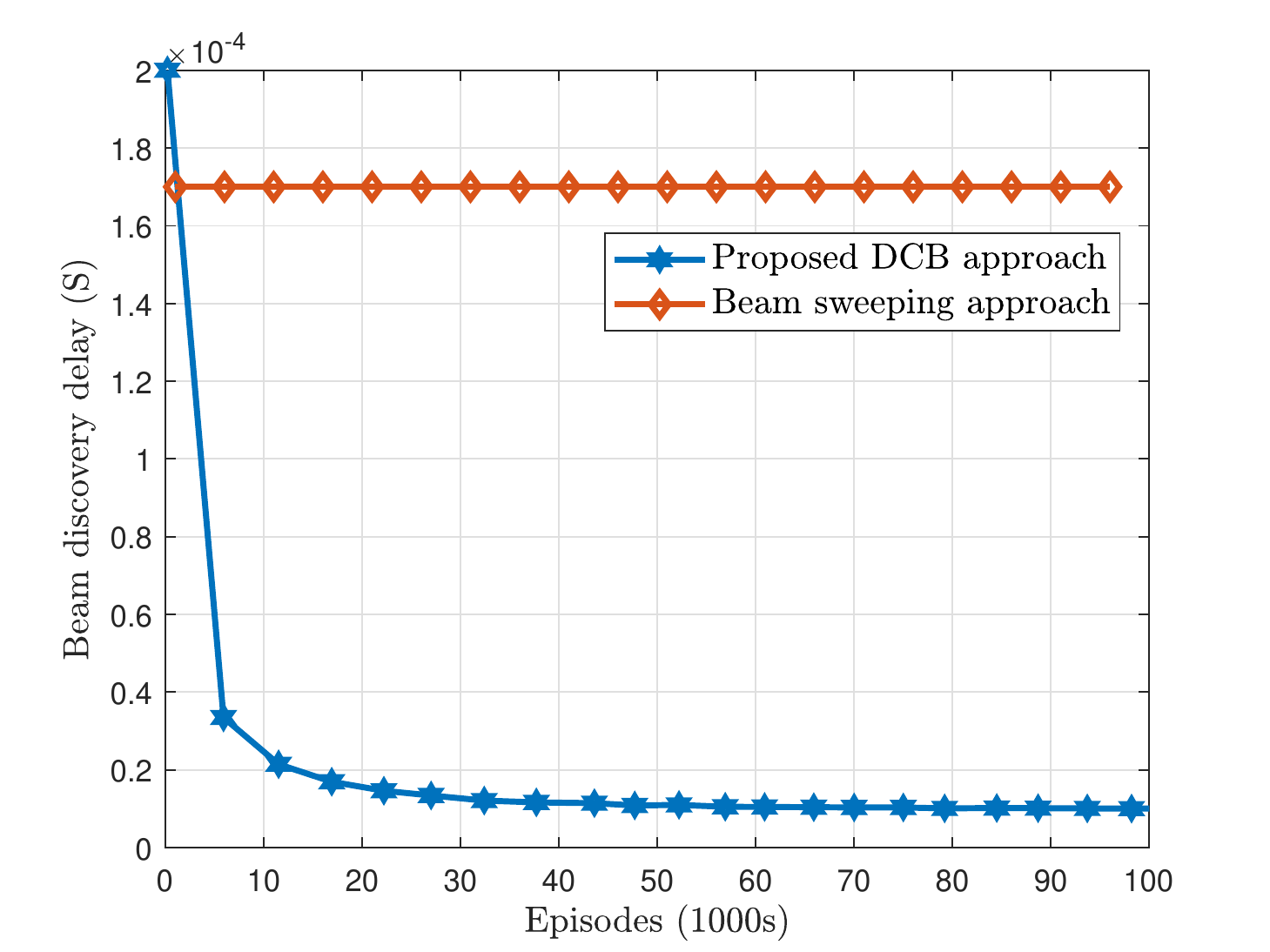}
	\caption{Average beam discovery delay, vs. training episodes for the ray-tracing based channel model.}
	\label{fig:delay}
\end{figure}

\subsubsection{Ray-tracing channel model}

All antenna gains and the noise figures are set to 5 and 3 dB, receptively.
An indoor sports stadium is chosen as the environment as shown in Fig. \ref{fig:simenvironment}.
It is an ideal candidate for early deployment and testing of mmWave UC UD networks due to a number of possible future application scenarios relating to live sports entertainment. 
The environment is modeled using Wireless Insite ray-tracer\cite{REMCOM2017}.

A 60 m $\times$ 30 m skating rink is situated in the middle of the 100 m $\times$ 50 m stadium.
All walls, ceiling, and floor are concrete structures and they are modeled using ITU 28 GHz compliant material models\cite{REMCOM2017}.
The perimeter walls around the stadium extend from the floor to the ceiling.
The seating area has an elevation angle of 30 degrees.
APs are located 2 m apart facing down in a grid formation on the ceiling which is 15 m above the floor.
A gird of 20 m $\times$ 10 m with a resolution of 0.25 m on the seating area is defined as the set of possible user locations. 
AP and possible user locations are shown in green and red color cubes in Fig. \ref{fig:simenvironment}.
In each simulation, 30 users are randomly placed on the grid.

\subsubsection{Stochastic channel model}
A simulation area of 200 m $\times$ 200 m is considered.
The number of APs in a simulation iteration depends on the AP density which is set between 40 and 200 APs per km$^2$.
At each iteration, 30 users are uniformly placed in the simulation area.
The channel parameters for path loss, shadowing, angular dispersion, and spatial clusters are selected based on \cite{Akdeniz2014}.

\subsubsection{DCB model}
%The DNN in the DCB model is trained for a $10^ 6$ iterations.
The DCB agent is implemented using the Keras\cite{keras} library.
DNN model of the DCB agent has 4 hidden layers where the Relu activation function and Adam optimizer are used.
The first three hidden layers have 50 neurons each while the last hidden layer has 16 neurons.
Input and output layers contain 32 and 16 neurons, respectively, corresponding to the size of a context and the action-space.
Past experiences are held in a first-in-first-out queue of length 50,000.
The values for $\epsilon_{dec}$ and $\epsilon_{min}$ are set to 0.99995 and 0.01, respectively.
Initially, 10,000 random guesses are made to explore the environment.
After this stage, an $\epsilon$-greedy policy is followed where $\epsilon$ decays by a factor of $\epsilon_{dec}$.
Experience-replay uses 64 randomly selected experiences from the queue.

\begin{figure}[t]
	\centering
	\includegraphics[width=\linewidth]{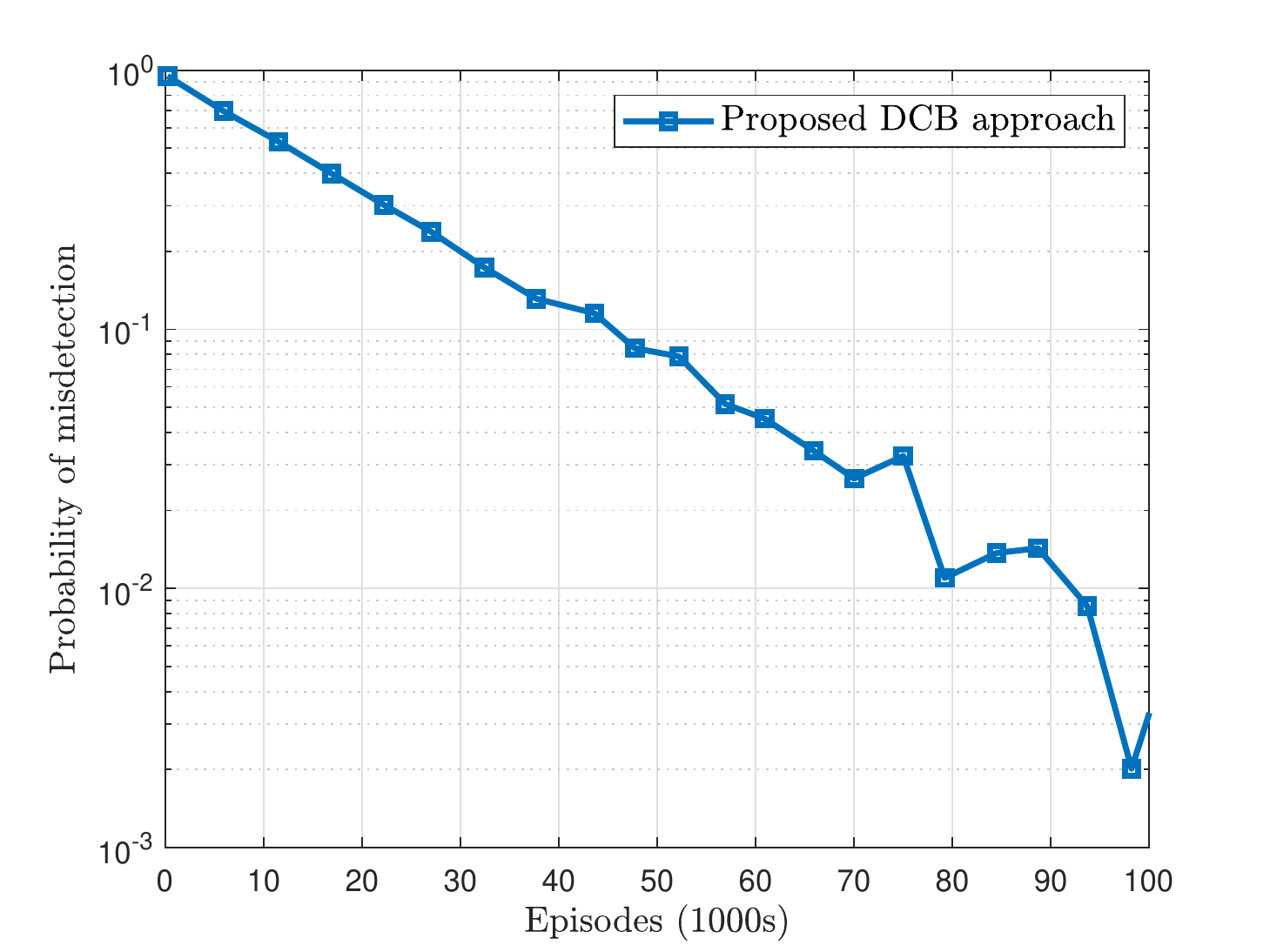}
	\caption{Probability of beam misdetection for the proposed approach, against the training episodes, for the ray-tracing based channel model.}
	\label{fig:p_misVsEpisodes}
\end{figure}

\subsection{Results}
\label{sec:results} 
Fig. \ref{fig:delay}, and \ref{fig:p_misVsEpisodes} show beam discovery delay, and the probability of misdetection for the indoor sports stadium simulation setting where the channel is modeled using ray-tracing.
An ultra-dense setting of 200 APs per km$^2$ is considered.
Fig. \ref{fig:delay} depicts the key advantage of the proposed approach--the low beam discovery delay.
In addition to the proposed approach, an interference-free case for the beam sweeping mechanism is considered.
The proposed DCB based approach is able to maintain a very small average beam discovery delay compared to the beam sweeping approach; 
the proposed approach uses only one reference signal from the user for the beam prediction.
The best case for beam sweeping, i.e., an interference-free scenario, has to transmit 16 beams and receive a report from the user to choose the IA beam for the user.
Hence, the beam sweeping method incurs a beam discovery delay of at least 0.17 ms.

Fig. \ref{fig:p_misVsEpisodes} illustrates how the proposed approach improves its prediction capability over time. 
The DCB model in the proposed approach can reach well below 10$^{-2}$ for the probability of beam misdetection even under ultra-dense conditions. 
Ray tracing based realistic simulations show that the proposed approach can support IA in mmWave based UC UD networks while maintaining satisfactory beam discovery delay, and the probability of beam misdetection.

\begin{figure}[t]
 	\centering
 	\includegraphics[width=\linewidth]{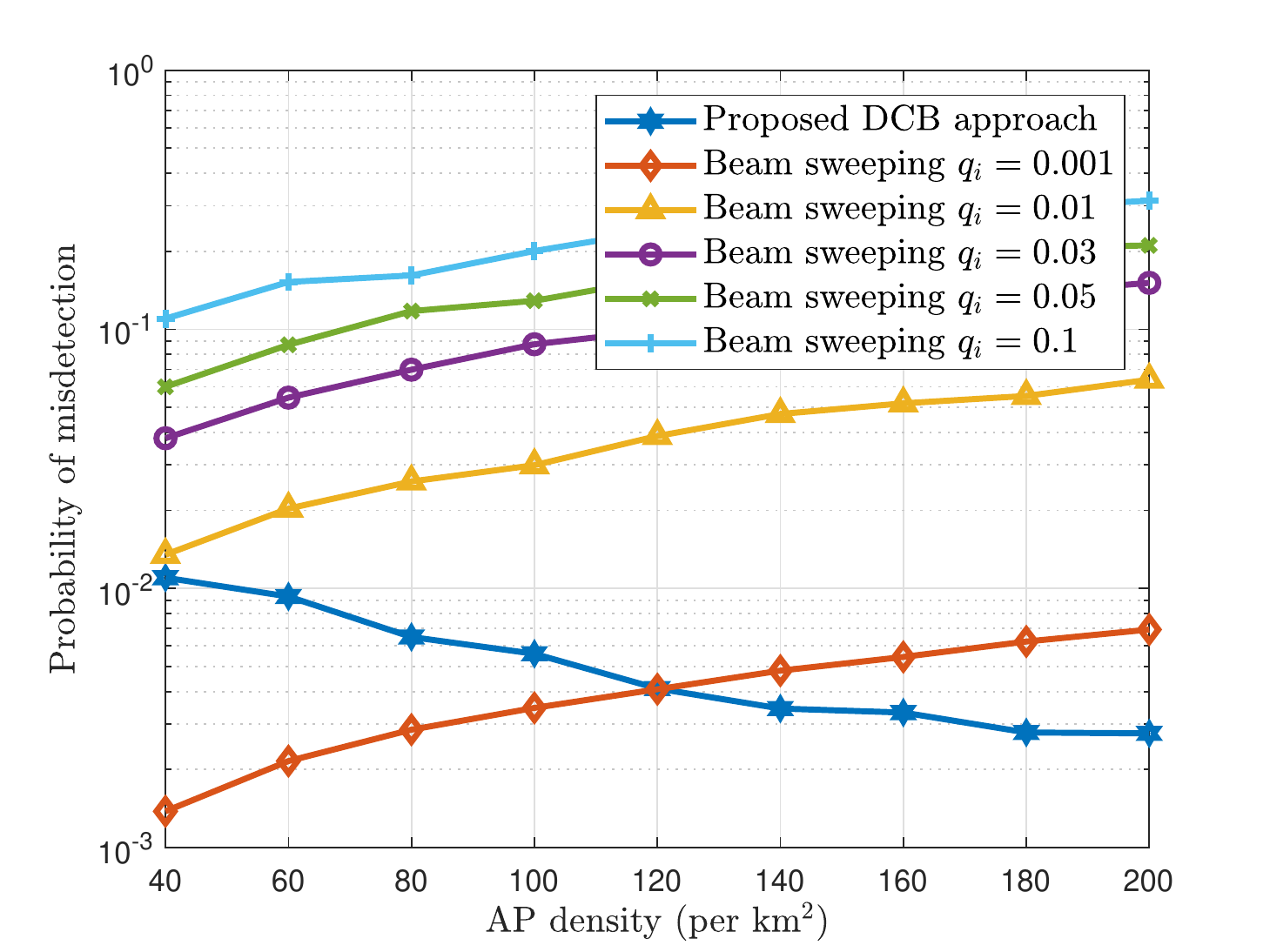}
 	\caption{Probability of beam misdetection, vs. the AP density, for the stochastic channel model.}
 	\label{fig:p_misVsDensity}
 \end{figure}

Fig. \ref{fig:p_misVsDensity} demonstrates how the effects of beam interference in the mmWave based UC UD networks dominate the probability of misdetection performance of the beam sweeping based approaches. 
We consider scenarios from almost no interference, i.e., $q_i=0.001$, to some interference, i.e., $q_i=0.1$.
The proposed approach remains unaffected by such beam interference and continues to show increase performance with the density.
For the low-density case of the proposed approach, the probability of the reference signal from the user not reach the sparsely located APs is high due to the characteristic of mmWave propagation.
However, in the high-density case, the impact of this becomes less, and the proposed approach performs better than the best beam sweeping case, i.e., the scenario with almost no interference
 
The proposed approach exhibits a low average beam discovery delay while maintaining a satisfactory probability of beam misdetection, compared to beam sweeping based methods. 
This feat is achieved through the main difference in the proposed approach--the IA process is performed using just one reference signal from the user.
%Moreover, the proposed approach allows the use of large codebooks which was previously constrained due to the lack of orthogonal resources available for an AP.

	\section{Conclusion}
	\label{sec:conclution}
	In this paper, we have applied deep contextual bandits (DCB) to the initial access (IA) problem in millimeter wave (mmWave) based user-centric (UC) ultra-dense (UD) networks.
	First, we have studied beam sweeping based IA procedures used in current systems. 
	Furthermore, we identified two problems with the extension of these methods to the UD UC network setting: 
	high beam discovery delay, and lack of radio resources to perform beam sweeps using large codebooks.
	Then, we propose a DCB based solution to perform fast and efficient IA by reducing the message transfers between the access points and the users.
	The DCB model predicts the beam for IA using a single reference signal from the user.
	Simulations were performed using geometric ray-tracing and stochastic models to evaluate the performance. 
	The simulation results show that the proposed system can predict the IA beam with a small beam discovery delay compared to the beam sweeping method while maintaining a low probability of beam misdetection.
	The proposed approach outperforms beam sweeping systems in the user-centric ultra-dense networks.
	Hence, with the evolution of communication systems to mmWave based UC UD networks, the beam sweep based IA strategies will become obsolete--due to the higher beam discovery delay and misdetection probability caused in beam sweeping systems.
	Work in this paper can be extended to several research directions: investigating the performance of the proposed approach under low received SNR setting, its extension towards coordinated beamforming, and fine-tuning the hyper-parameters in the machine learning model.

%	\begin{figure}[t]
%		\centering
%		\includegraphics[width=0.975\linewidth]{../Figs/Results_07-06/delay}
%		\caption{Average beam discovery delay for a user using RapidIA, DeepIA-DRL and BSIA algorithms.}
%		\label{fig:delay}
%	\end{figure}

%\section*{Acknowledgment}
%The authors thank Mr. Tarun Chawla from Remcom for extending technical support and expertise for this work.
%
%This work has been financially supported in part by the
%6Genesis (6G) Flagship project (grant 318927)
	
\bibliographystyle{IEEEtran}
\bibliography{MSc-Paper3} 
	
\end{document}